\documentclass[prl,twocolumn]{revtex4-1}
\usepackage[utf8x]{inputenc}
\usepackage{amsbsy}
\usepackage{graphicx}
\usepackage{color}
\usepackage{dcolumn}
\usepackage{amsmath}
\usepackage{amssymb}
\usepackage{amsfonts}
\usepackage{subfigure}
\usepackage{bm}

\usepackage[colorlinks=true, citecolor=black, linkcolor=black, urlcolor=black]{hyperref}

\begin{document}
\title{Hydrogen sulphide at high pressure: a strongly-anharmonic
  phonon-mediated superconductor}

\author{Ion Errea$^{1,2}$} 
\author{Matteo Calandra$^{3}$} \email[]{matteo.calandra@upmc.fr}
\author{Chris J.\ Pickard$^{4}$}
\author{Joseph Nelson$^5$}
\author{Richard J.\ Needs$^5$} 
\author{Yinwei Li$^6$}
\author{Hanyu Liu$^7$}
\author{Yunwei Zhang$^{8}$}
\author{Yanming Ma$^8$} 
\author{Francesco Mauri$^{2}$} 

\affiliation{$^1$Donostia International Physics Center
(DIPC), Manuel de Lardizabal pasealekua 4, 20018 Donostia-San
Sebasti$\tilde{a}$n, Basque Country, Spain} 
\affiliation{$^2$IKERBASQUE, Basque
Foundation for Science, Bilbao, Spain}
\affiliation{$^3$IMPMC, UMR CNRS 7590, Sorbonne
Universit\'es - UPMC Univ. Paris 06, MNHN, IRD, 4 Place Jussieu,
F-75005 Paris, France}
\affiliation{$^4$Department of Physics \&
Astronomy, University College London, Gower Street, London WC1E~6BT,
UK} 
\affiliation{$^5$Theory of Condensed Matter
Group, Cavendish Laboratory, J J Thomson Avenue, Cambridge CB3 0HE,
UK} 
\affiliation{$^6$School of Physics and Electronic
Engineering, Jiangsu Normal University, Xuzhou 221116, People's
Republic of China} 
\affiliation{$^7$Department of Physics and Engineering
Physics, University of Saskatchewan, Saskatchewan S7N 5E2, Canada}
\affiliation{$^8$State Key Laboratory of Superhard
Materials, Jilin University, Changchun 130012, People's Republic of
China} 

\date{\today}

\begin{abstract}
  We use first principles calculations to study structural,
  vibrational and superconducting properties of H$_2$S at pressures $P
  \ge 200$ GPa. The inclusion of zero point energy leads to two
  different possible dissociations of H$_2$S, namely 3H$_2$S $\to$
  2H$_3$S + S and 5H$_2$S $\to$ 3H$_3$S + HS$_2$, where both H$_3$S
  and HS$_2$ are metallic.  For H$_3$S, we perform non-perturbative
  calculations of anharmonic effects within the self-consistent
  harmonic approximation and show that the harmonic approximation
  strongly overestimates the electron-phonon interaction
  ($\lambda\approx 2.64$ at 200 GPa) and T$_c$.  Anharmonicity hardens
  H--S bond-stretching modes and softens H--S bond-bending modes.  As
  a result, the electron-phonon coupling is suppressed by $30\%$
  ($\lambda\approx 1.84$ at 200 GPa).  Moreover, while at the harmonic
  level T$_c$ decreases with increasing pressure, the inclusion of
  anharmonicity leads to a T$_c$ that is almost independent of
  pressure.  High pressure hydrogen sulfide is a strongly anharmonic
  superconductor.
\end{abstract}

\maketitle

Cuprates \cite{Bednorz} have for many years held the world record for
the highest superconducting critical temperature (T$_c=133$ K)
\cite{Schilling}.  However, despite almost $30$ years of intensive
research, the physical mechanism responsible for such a high T$_c$ is
still elusive, although the general consensus is that it is highly
non-conventional.  The discovery by Drozdov \textit{et al.}\
\cite{Drozdov} of T$_c=190$ K in a diamond anvil cell loaded with
hydrogen sulfide (H$_2$S) and compressed to about 200 GPa breaks the
cuprates record and overturns the conventional wisdom that such a high
T$_c$ cannot be obtained via phonon-mediated pairing.

The claim that hydrogen at high pressure could be superconducting is
not new \cite{Ashcroft} and it was recently supported by first
principles calculations based on the harmonic approximation applied to
dense hydrogen \cite{Ma, Cudazzo2008, Cudazzo2010a, Cudazzo2010b} and
several hydrides \cite{Kim, Scheler, Zhou, KimPNAS, Gao, GaoPRL,
  Feng_SiH4_2006}.  More recently, two theoretical papers predicted
the occurrence of high T$_c$ superconductivity in high-pressure
sulfur-hydrides \cite{Duan, Y_Li}.  However, as shown in Refs.\
\cite{ErreaPRL, ErreaPtH}, anharmonicity can be crucial in these
systems. For example, in PdH, the electron-phonon coupling $\lambda$
parameter is found to be $1.55$ at the harmonic level, while a proper
inclusion of anharmonic effects leads to $\lambda=0.40$
\cite{ErreaPRL}, in better agreement with experiments.  Thus, in
hydrogen-based compounds, the phonon spectra are strongly affected by
anharmonic effects.


Several first principles calculations \cite{Duan,Y_Li, Flores-Livas,Arita}
suggested that decomposition of the H$_2$S sample occurs within the
diamond-anvil cell at high pressures.  The high-T$_c$ superconducting
material is therefore very unlikely to be H$_2$S, while H$_3$S is the
obvious candidate for the H-rich decomposition product.

Here we study the structural, vibrational and superconducting
properties of H$_2$S above $200$ GPa, where the highest T$_c$ occurs.
We show that the inclusion of zero point motion in the convex hull at
200 and 250 GPa stabilizes two metallic structures, H$_3$S and HS$_2$.
Finally, we show that, contrary to suggestions in previous work
\cite{Duan, Flores-Livas}, the harmonic approximation does not explain
the measured T$_c$ in H$_3$S, and the inclusion of anharmonic effects
is crucial.

As decomposition of H$_2$S has been demonstrated in the experiments of
Ref.\ \cite{Drozdov}, it is crucial to develop an understanding of the
different H/S compounds that might be stable in the pressure range of
interest. We therefore perform a search over 43 H/S stoichiometries,
determining the stoichiometries at which stable structures exist, and
the associated crystal structures. These searches were performed using
the \textit{ab initio} random structure searching (AIRSS) method
\cite{Pickard_silane_2006, AIRSS_review_2011} and the CALYPSO particle
swarm optimization method of Ref. \cite{Ma3}.
 More information about the searches
is provided in the Supplemental Material \cite{SI}.
\begin{figure}[t]
\includegraphics[width=0.9\columnwidth]{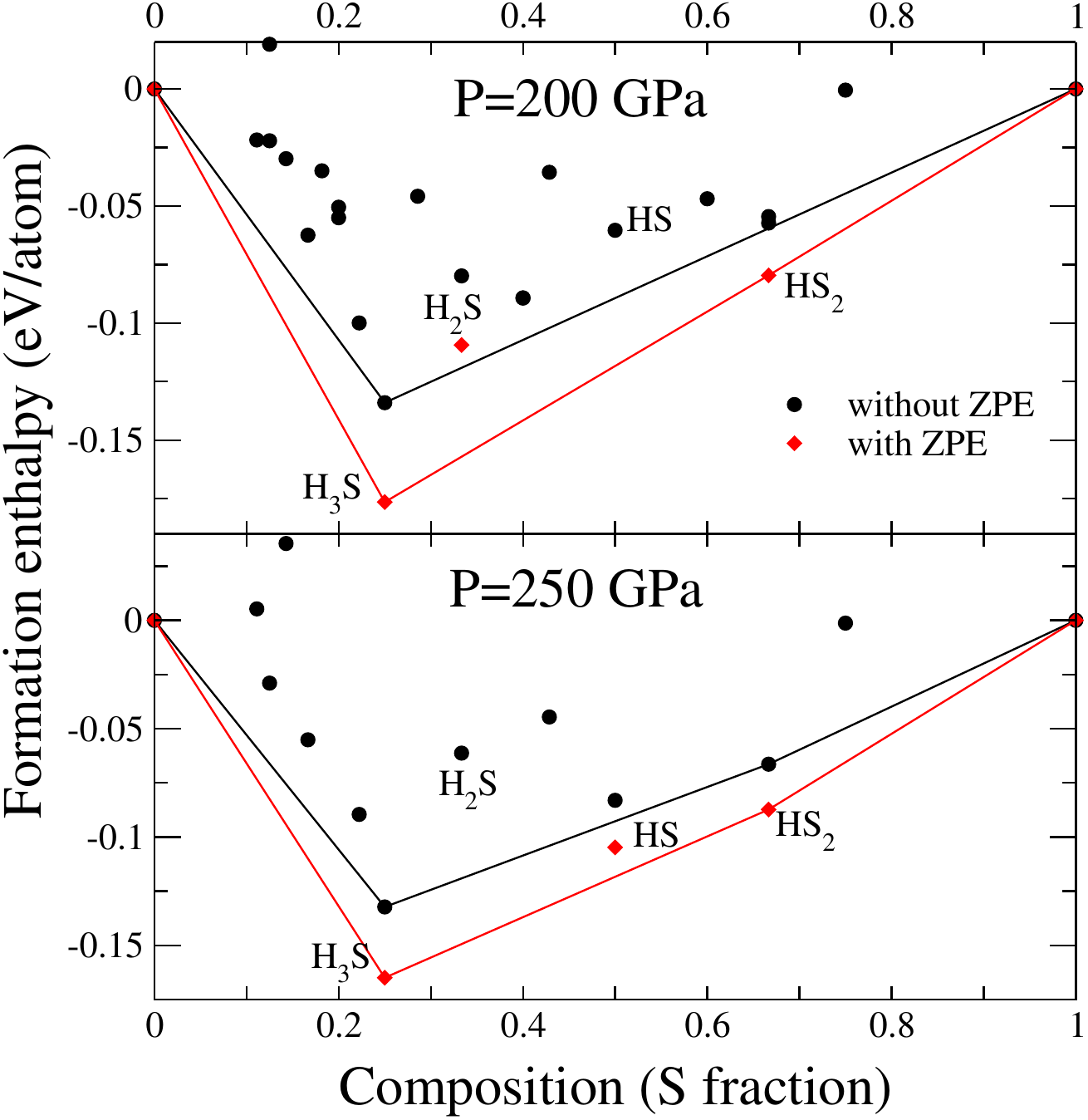}
\caption{Results of structure searching at 200 and 250 GPa. Convex
  hulls are shown as continuous lines, with and without the inclusion
  of zero point energy (ZPE).}
\label{fighulls}
\end{figure}

The results of the structure searching are shown in Fig.\
\ref{fighulls}. At 200 GPa, without zero-point energy (ZPE), the only
energetically allowed decomposition is 3H$_2$S $\to$ 2H$_3$S + S, in
agreement with previous calculations \cite{Duan, Bernstein,
  Flores-Livas, Arita}. H$_3$S crystallizes in the space group
$Im\overline{3}m$, as shown in \cite{Duan}.  When ZPE is included, a
second decomposition becomes possible at 200 GPa, namely 5H$_2$S $\to$
3H$_3$S + HS$_2$, where HS$_2$ crystallizes in a structure of space
group {\it C2/c} with $12$ atoms/cell. At 250 GPa and above, the
latter decomposition is allowed even without ZPE, and the {\it C2/c}
HS$_2$ structure undergoes a phase transition to a more stable {\it
  C2/m} structure with $6$ atoms/cell. Each of the HS$_2$ structures
is metallic. Finally, at 300 GPa, an HS phase becomes stable
\cite{SI}.  Detailed information on the crystal structures is provided
in the Supplemental Material \cite{SI}.

Having determined the most stable crystal structures at high pressure,
we turn to the study of vibrational properties \cite{QE, TecDet}.  We
consider the $Im\overline{3}m$ H$_3$S structure at 200 GPa
\cite{footnote}. In this structure each H atom is twofold coordinated
and has $6$ neighbors, $2$ of which are S atoms while the other $4$
are $H$ atoms.  H vibrations can then be decomposed into
HS bond-stretching modes (H$_\parallel$), in which an H atom moves
towards one of the two S atoms, and bond-bending modes (H$_\perp$), in
which one H atom moves in the direction perpendicular to the H--S bond
(see Fig.\ 1 in \cite{SI}).
\begin{figure}[t]
\includegraphics[width=0.9\columnwidth]{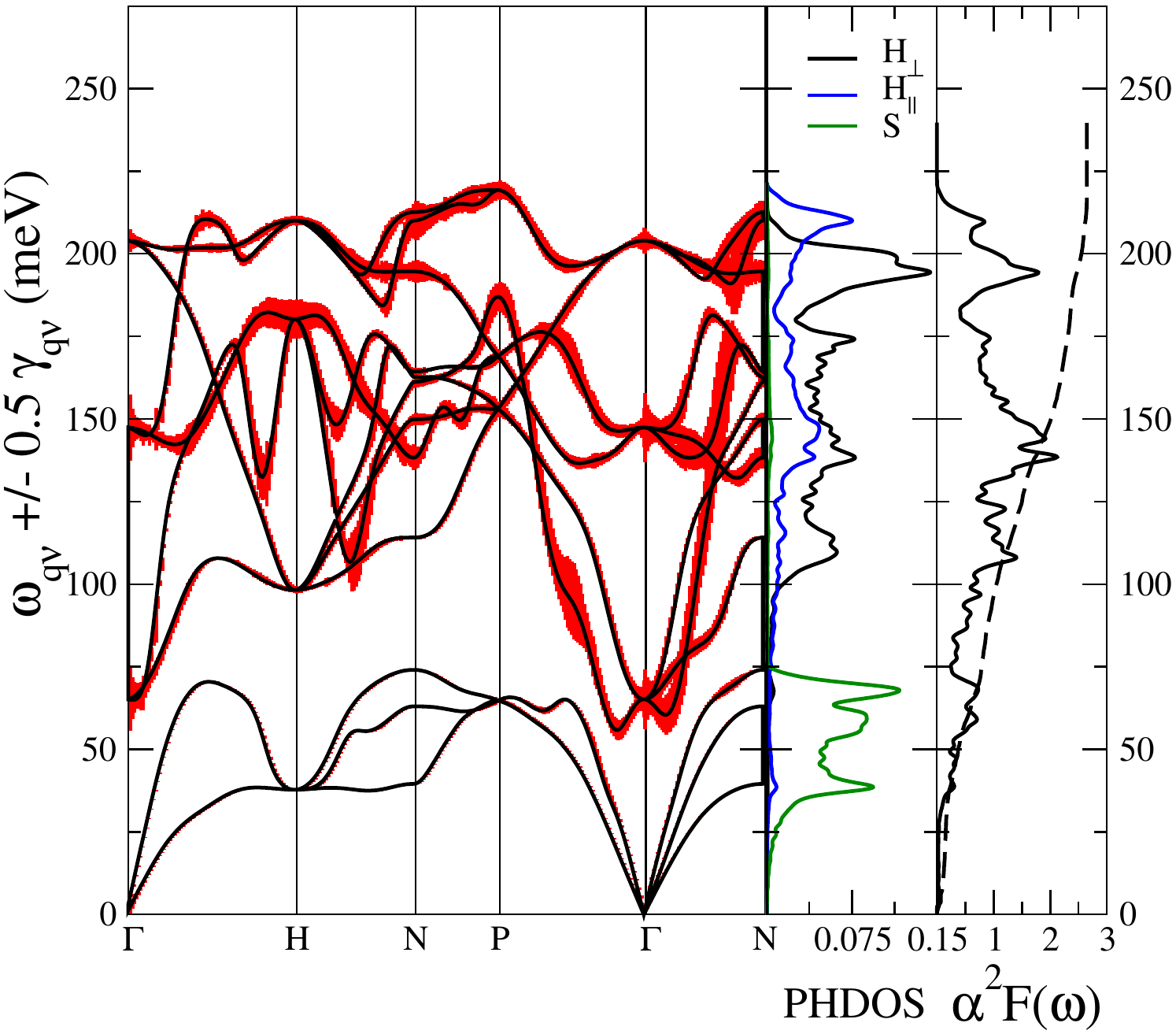}
\includegraphics[width=0.9\columnwidth]{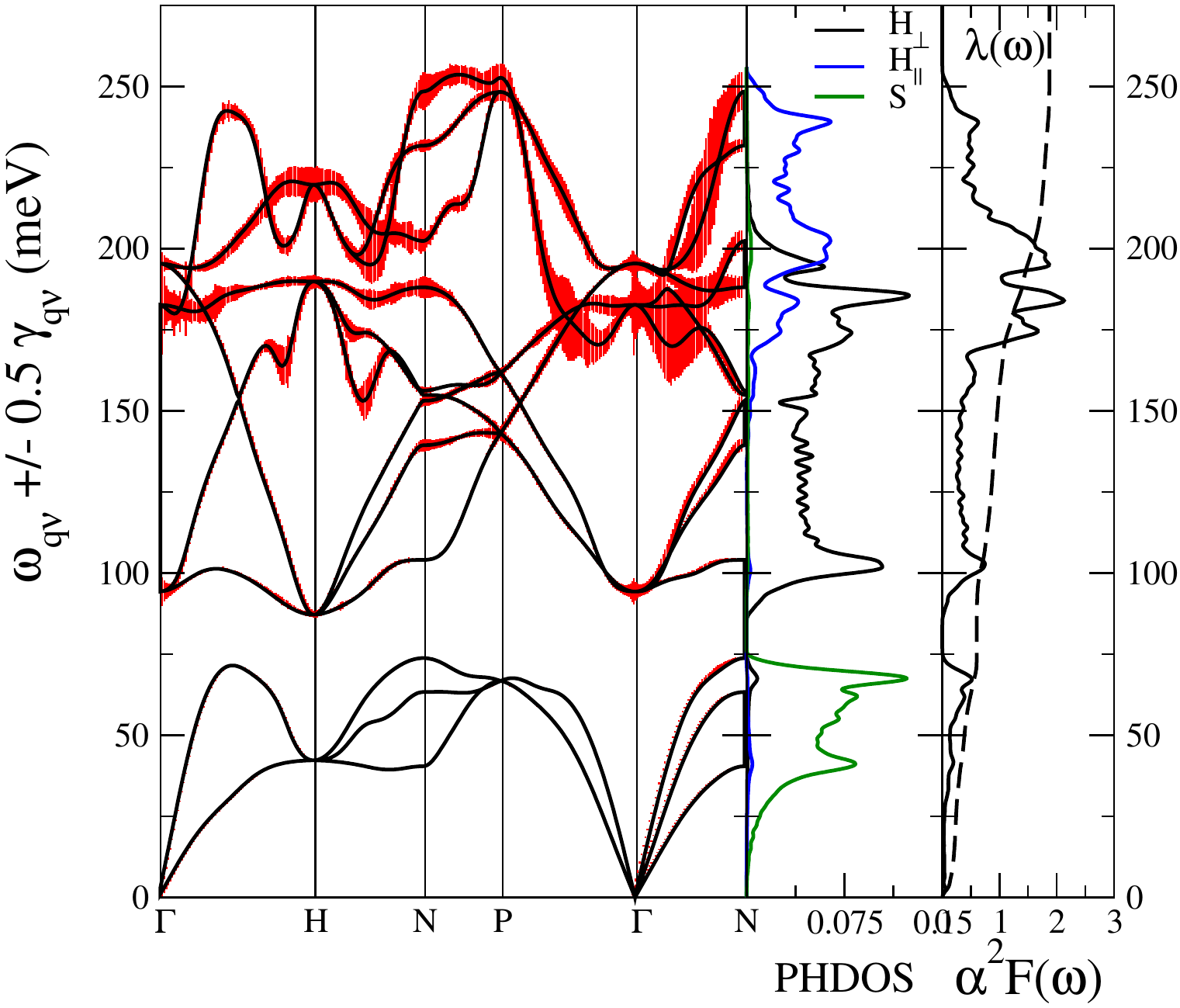}
\caption{Phonon dispersion, phonon density of states projected onto
  selected atoms and directions, and the Eliashberg function of H$_3$S
  in the harmonic approximation (top) and with the inclusion of
  anharmonic effects (bottom) for H$_3$S at 200 GPa. H$_\perp$ and
  H$_\parallel$ label displacements of an H atom in the directions
  perpendicular or parallel to a H--S bond. The magnitude of the
  phonon linewidth is indicated by the size of the red error bars.}
\label{figh3sharm}
\end{figure}
The harmonic phonon spectrum of H$_3$S is shown in Fig.\
\ref{figh3sharm} and overall shows a clear separation into H modes at
high energy and S modes below 75 meV.  To gain more insight we use
Wannier interpolation \cite{CalandraWannier, Giustino} of the
electron-phonon matrix elements and evaluate the electron-phonon
contribution to the phonon linewidth, as \cite{Allen}:
\begin{equation}
\label{eq:gamma_Allen}
\gamma_{{\bf q}\nu} = \frac{4\pi \omega_{{{\bf q} \nu}}}{N_k}
\sum_{{\bf k},n,m} 
|g_{nm}^{\nu}({\bf k}, {\bf k}+ {\bf q})|^2 \delta(\varepsilon_{{\bf k}n}) \delta(\varepsilon_{{\bf k+q}m}) \ . 
\end{equation}
Here $\omega_{{{\bf q} \nu}}$ are the phonon frequencies, $N_k$ the
number of electron-momentum points in the grid, $g_{nm}^{\nu}({\bf k},
{\bf k} + {\bf q})= \langle {\bf k} n|\delta V_{KS}/\delta u_{{\bf
    q}\nu} |{\bf k+q}m \rangle $ is the electron-phonon matrix
element, $V_{KS}$ is the Kohn-Sham potential, and $u_{{\bf q}\nu} $ is
a phonon displacement. The Kohn-Sham energy and eigenfunctions are
labeled $\varepsilon_{{\bf k}n}$ and $|{\bf k}n\rangle$.  The
electron-phonon coupling at a given phonon-momentum ${\bf q}$ for a
phonon mode $\nu$ can be obtained \cite{Allen} from the phonon
linewidth as $\lambda_{{\bf q}\nu}= \frac{{\gamma}_{{\bf q}\nu}}{2\pi
  \omega_{{\bf q}\nu}^2 N(0)}$.
 
As shown in Fig.\ \ref{figh3sharm}, at the harmonic level, the phonon
linewidths of the H vibrations is fairly uniform throughout the
spectrum.  The contribution of each mode to the average
electron-phonon interaction, $\lambda=\sum_{\nu {\bf q}}\lambda_ {{\bf
    q}\nu}/N_q$, can be obtained from the isotropic Eliashberg
function
\begin{equation}
  \alpha^2F(\omega)=\frac{1}{2 N_{q}}\sum_{{\bf q}\nu} 
  \lambda_{{\bf q}\nu} \omega_{{\bf q}\nu} \delta(\omega-\omega_{{\bf q}\nu} )
\label{eq:Eliashberg_function}
\end{equation}
where $N_{q}$ is the number of phonon-momentum points in the grid.
$\lambda(\omega)=2\int_{0}^{\omega} \frac{\alpha^2F(\omega')}{\omega'}
\,d\omega'$ and then $\lambda=\lambda(\infty)$. We find $\lambda=2.64$
(see Table \ref{table1}), which is larger than that obtained in Refs.\
\cite{Duan, Flores-Livas, Arita} with a much coarser sampling of the
BZ.  This huge value of $\lambda$ comprises substantial contributions
from many H vibrational modes. The situation is therefore very
different from MgB$_2$ in which a single mode dominates $\lambda$.

Given the low mass of H and the consequent large phonon displacements,
we investigate the occurrence of anharmonic effects using the
stochastic self-consistent harmonic approximation (SSCHA) developed by
some of us \cite{ErreaPRL, ErreaPtH, SSCHAtechdet}.  As shown in Fig.\
\ref{figh3sharm} (bottom), the anharmonic correction leads to
non-trivial changes in the harmonic spectrum. While it is very clear
that all H bond-stretching modes are hardened, the effect on H
bond-bending modes is less straightforward.  By computing the average
phonon frequency as of H$_{\parallel}$ and of H$_{\perp}$ modes we
find that $\overline{\omega}_{\parallel}^{\rm har}\approx 158.1$ meV
and $\overline{\omega}_{\parallel}^{\rm anh}\approx 203.3$ meV, while
for bond-bending modes $\overline{\omega}_{\perp}^{\rm har}\approx
157.0$ meV and $\overline{\omega}_{\perp}^{\rm anh}\approx 147.9$ meV.
Thus bond-stretching modes are hardened, while bond-bending modes are
softened.

It is important to remark that the large and most dispersive mode
along P$\Gamma$ is strongly hardened at the anharmonic level and
undergoes a non-trivial change in polarization, as can be seen from
the large effect of anharmonicity on the phonon linewidth
$\gamma_{{\bf q}\nu}$ in Fig.\ \ref{figh3sharm}.  It is worhwhile to
recall that the phonon-linewidth depends on the phonon eigenvector but
not on the phonon energy.  This effect demonstrates the need to
calculate not only the phonon frequencies at the anharmonic level, but
also the phonon polarizations.

The anharmonic electron-phonon interaction is $\lambda=1.84$, which is
$30\%$ smaller than the harmonic result.  This reduction is mostly
explained by the hardening of the H$_{\parallel}$ modes.  In contrast
to the harmonic case which shows uniform coupling over all modes, the
anharmonic Eliashberg function has two main peaks, a broad peak in the
40--75 meV region, and a second one in the 175--200 meV region.  Their
contributions to $\lambda$ are $0.59$ and $0.77$, respectively,
accounting for $\approx 73\%$ of the total $\lambda$.  We note,
however, that the logarithmic average of the phonon frequencies,
$\omega_{\log}$, is only weakly enhanced by anharmonicity (see Table
\ref{table1}).

The superconducting critical temperature can be obtained either from
the McMillan equation or the isotropic Migdal-Eliashberg
approach. However, it is well known \cite{AllenDynes} that the use of
the McMillan equation for such values of $\lambda$ leads to a
substantial underestimation of T$_c$. We solved the isotropic
Eliashberg equations \cite{SI} and found, contrary to claims in
previous publications \cite{Duan, Flores-Livas}, that
calculations based on the harmonic phonon spectrum do not explain the
measured T$_c$ as, even using large values of $\mu^{*}$ \cite{Morel,
  Bogolubov}, T$_c$ is substantially overestimated (i.e., T$_c=250$ K
for $\mu^{*}=0.16$ \cite{SI}).  When the anharmonic phonon spectrum
and electron-phonon coupling are used, the Migdal-Eliashberg equations
account for the experimental T$_c$ when the value $\mu^*=0.16$ is
used, as shown in Table \ref{table1}.  The superconducting gap at zero
temperature is $\Delta\approx 36.5$ meV.

Interestingly, the large anharmonic effects lead to very different
variation of T$_c$ with pressure. By repeating the calculation for the
$Im\overline{3}m$ structure at 250 GPa, we found at the harmonic level
and using the Migdal-Eliashberg equations with the same values of
$\mu^*=0.16$, that T$_c=226$ K, decreasing with increasing
pressure. However, at the anharmonic level we find T$_c=190$ K,
essentially independent of pressure in the region 200--250 GPa.

Finally, we consider the extent to which the occurrence of large
anharmonic effects can explain the isotope shift in D$_2$S. At $164$
GPa, T$_c({\rm D_2S})=90$ K, leading to an isotope coefficient
$\alpha\approx 1.07$, which is substantially enhanced from the
canonical BCS value of $\alpha\approx 0.5$.  Assuming a similar
decomposition of D$_2$S into D$_3$S and S at high pressures, we
calculate the anharmonic phonon spectrum (see \cite{SI}) and
electron-phonon coupling in D$_3$S at $200$ GPa.  We find at the
anharmonic level that the electron-phonon coupling is essentially
unaffected, while $\omega_{\log}$ is softened from $92.9$ meV to
$73.3$ meV, leading to an isotope coefficient of $\alpha=0.35$, which
is strongly reduced from the BCS value but inconsistent with the value
of $\alpha\approx 1.07$ found in experiments. Thus, contrary of what
claimed in Ref. \onlinecite{Papaconstantopoulos}, anharmonicity
reduces $\alpha$.
\begingroup
\squeezetable
\begin{table}[h]
  \caption{Electron-phonon interaction and logarithmic averages of 
    phonon frequencies, with and without anharmonic effects. The
    T$_c$s are calculated using the isotropic Migdal-Eliashberg 
    equations (T$_c^{\rm ME}$). A value of $\mu^*=0.16$ is used. 
    Data for T$_c$ calculated with the McMillan equation is provided 
    in the Supplemental Material \cite{SI}. Frequencies are in meV and
    T$_c$s are in K.} 
\begin{tabular}{l c c c c c c c } 
\hline
\hline
Compound &  $\lambda^{\rm har.} $ & $\omega_{\rm log}^{\rm har}$ &
$\lambda^{\rm anh}$ & $\omega_{\log}^{\rm anh}$ & T$_c^{\rm ME,
  har.}$  &  T$_c^{\rm ME, anh}$ & T$_c$(Exp)\\
\hline
H$_3$S (200 GPa) & 2.64 & 90.4  & 1.84  & 92.86 &  250 &
194.0 &  190\\
H$_3$S (250 GPa) & 1.96 & 109.1  & 1.71  & 101.3 &  226 & 190 &  \\
D$_3$S (200 GPa) & 2.64 & 68.5 &  1.87  & 73.3  & 183 &
152.0 &  90\\\hline
\end{tabular}
\label{table1}
\end{table}
\endgroup

We have studied the structural, vibrational and superconducting
properties of high pressure H$_3$S.  We have included zero point
motion when comparing the stabilities of different H/S phases, which
has been neglected in other publications so far.  This is important
because zero point motion stabilises a new phase at $P\ge 200$ GPa.
In particular, we found that decomposition of HS$_2$ into metallic
phases can occur following two main paths, namely 3H$_2$S $\to$
2H$_3$S + S and 5H$_2$S $\to$ 3H$_3$S + HS$_2$.  We have performed a
detailed study of the vibrational properties of high pressure H$_3$S
and D$_3$S, finding that the phonon spectra are strongly affected by
anharmonic effects.  Anharmonicity hardens H--S bond-stretching modes
and softens H--S bond-bending modes. Moreover, anharmonicity leads to
a reduction in the electron-phonon coupling by $\approx 30\%$ and to
an approximately constant T$_c$ in the range 200--250 GPa.  Our work
demonstrates that the superconducting properties of high pressure
H$_3$S can only be properly described by including both nuclear
quantum effects and anharmonicity.

%
%

We acknowledge discussions with I.\ I.\ Mazin and support from the
Graphene Flagship and Agence nationale de la Recherche, grant n.\
ANR-13-IS10-0003-01.  Computer facilities were provided by PRACE,
CINES, CCRT and IDRIS.  I.E.\ acknowledges financial support from the
Department of Education, Language Policy and Culture of the Basque
Government (Grant No.\ BFI-2011-65) and the Spanish Ministry of
Economy and Competitiveness (FIS2013-48286-C2-2-P). C.J.P.\ and
R.J.N.\ thank EPSRC (UK) for financial support.  J.R.N.\ acknowledges
financial support from the Cambridge Commonwealth Trust.  Y.\ Li
thanks the National Natural Science Foundation of China under Grant
Nos.\ 11204111 and 11404148.  Y.\ Zhang and Y.\ Ma thank the Natural
Science Foundation of China under No.\ 11274136, the 2012 Changjiang
Scholars Program of China.


\clearpage
\large
\onecolumngrid
\subsection{Supplementary Materials of \\
{\it Hydrogen Sulfides at high-pressure: a strongly-anharmonic phonon-mediated superconductor.  }}

\subsection{Technical details for the structural searching}

We use the CALYPSO and AIRSS codes.
Structure searching were performed at 200, 250, and 300 GPa for
several H-S compounds 
(H8S, H7S, H6S, H5S, H4S, H3S, H2S, HS, H3S2, H2S3, H5S2, H9S2, HS2,
H4S3, HS3, H3S2, H7S2) 
with maximum eight formula unit in the models. Each generation
contained 
40 structures, and the first generation was produced randomly with
symmetry constraint. 
All structures were locally optimized using density functional theory
with the Perdew-Burke-Ernzerhof (PBE) \cite{PBE}.
generalized gradient approximation implemented in the Vienna ab initio
simulation package\cite{VASP1,VASP2}. 
An energy cutoff of 700 eV and a Monkhorst-Pack Brillouin zone
sampling 
grid with a resolution of 0.5 \AA$^{-1}$ were used in structure searches. 
The 60$\%$ lowest-enthalpy structures of each generation were used to 
produce the structures in the next generation by local PSO technique,
and the remaining 40$\%$ 
structures were randomly generated within symmetry constraint to
enhance 
the structural diversity. Typically, the structure searching
simulation 
for each composition was stopped when ~1000 successive structures 
were generated after a lowest energy structure was found. 
A number of distinct low-enthalpy structures found were then 
re-optimized with denser grids better than 0.2 \AA$^{-1}$ and a higher 
energy cutoff of 1000 eV. The lowest-enthalpy structures were 
then chose to draw the convex-hull. 
\clearpage

\subsection{Convex hull at 300 GPa}

\begin{figure}[h]
\includegraphics[width=0.45\textwidth]{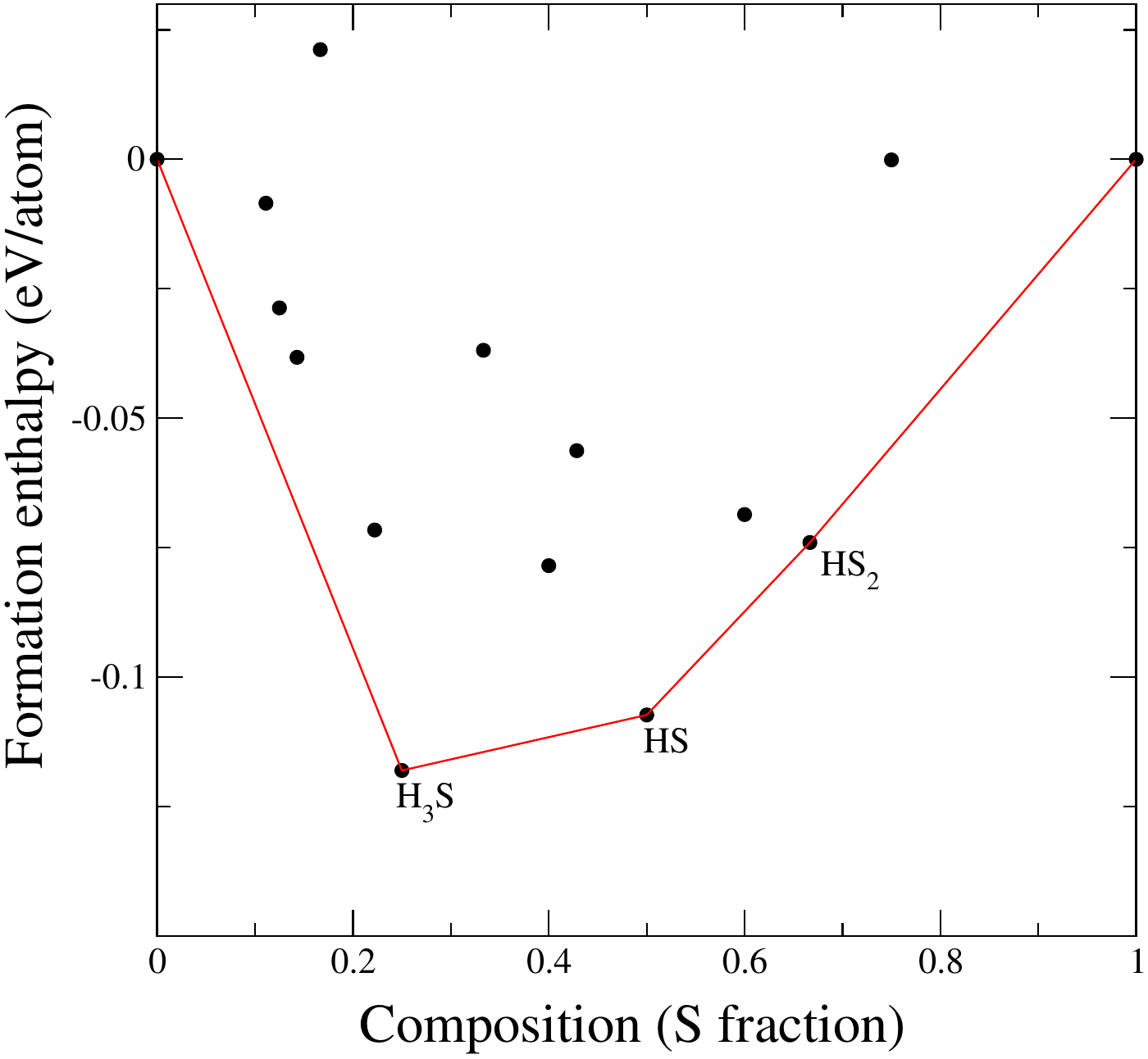}
\caption{Results of structural searches at 300 GPa. The continuous
  line shows the convex hull.}
\label{Hull_300}
\end{figure}

\subsection{Effect of the zero point energy on the D$_2$S convex hull
  at 200 GPa}

\begin{figure}[h!]
\includegraphics[width=0.45\textwidth]{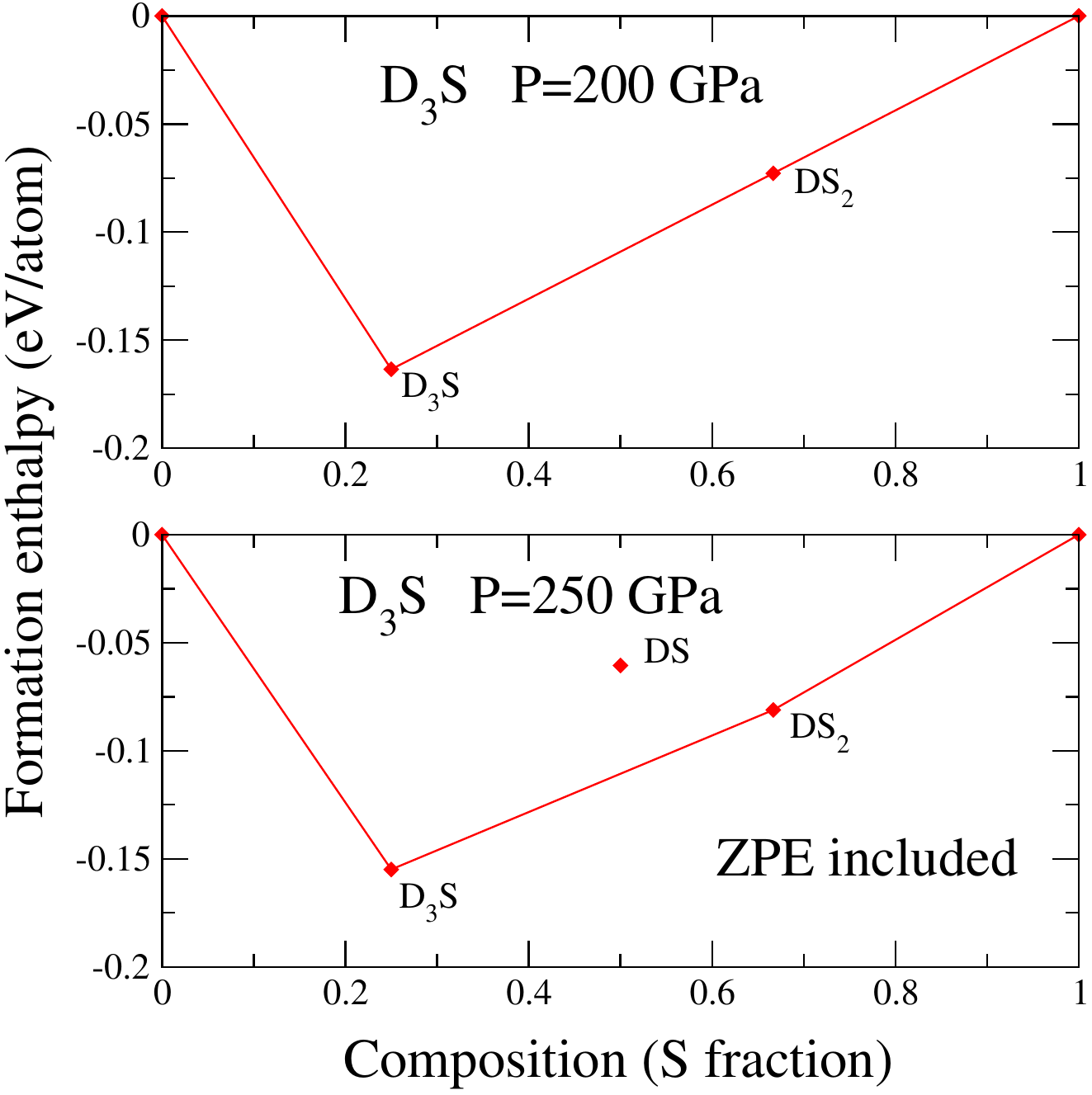}
\caption{Results of structural searches at 200 and 250 GPa for D-S
  structures including zero point energy. The continuous lines show
  the convex hull}
\label{Hull_DS}
\end{figure}

\clearpage

\subsection{Crystal structures for H and S}

The crystal structure for Hydrogen is from Ref.\ \cite{ChrisNatPhys}.
The crystal structure for S is from Ref. \cite{SStruc}

\subsection{Crystal structures of HS$_2$ at 200, 250 and 300 GPa}

\begin{table}[h]
  \caption{Crystallographic data for HS$_2$ at 200, 250 GPa as
    obtained from structural searches. At 250 and 300 GPa, HS$_2$ adopts
    the same C2/m structure as at $250$ GPa.}
\begin{center}
\begin{tabular*}{\textwidth}{l c c c c c c c}
\hline\hline
Structure & Pressure & Structural parameters & & & Atomic Positions &
& \\
               &     (GPa)  & (\AA, deg.)  & & & & & \\ \hline
C2/c       &  200  & a=6.7827, b=4.1876, c=7.5401 &  S  & 8f & 0.07702	&
0.12708& 	0.42711\\
              &           &$\alpha=90$, $\beta=137.7464$, $\gamma=90$ &
S & 8f & 0.89224	& 0.37399 &	0.79574\\
       &     &     & H & 8f &0.28345	&  0.86882	&  0.42463\\
C2/m    & 250 &  a=7.2073, b=2.947, c=3.6324, & S& 4i &
0.33820 &	0.5& 	0.51566\\
            &       & $\alpha=90$, $\beta=60.2287$,$\gamma=90$ &
S	& 4i &	0.59068& 	0.0& 	0.14405\\
            &   &    &   H &	4i& 	0.61860& 	0.5&
            0.94147\\\hline
\end{tabular*}
\end{center}
\label{tabHS2}
\end{table}

The HS$_2$ crystal structures found with the CALYPSO and AIRSS codes
are shown in Table \ref{tabHS2}. 

\clearpage
\subsection{Crystal structures of HS and at 200, 250 and 300 GPa}

\begin{table}[h]
  \caption{Crystallographic data for HS at 200 and 300 GPa as
    obtained from structural searches. At 250 GPa and 300 GPa, HS$_2$ adopts
    the same C2/m structure.}
\begin{center}
\begin{tabular*}{\textwidth}{l c c c c c c c}
\hline\hline
Structure & Pressure & Structural parameters & & & Atomic Positions &
& \\
I4$_1$/amd & 200 & a=b=2.9399, c=9.0531 &  S  & 16h & 0.0 & 0.0 &
0.27894 \\
                   &        &       &            H & 16h & 0.5 &0.0 & 0.62786\\
C2/m    & 300 & a=9.4579, b=2.7388, c=2.749, & S &	4i &
0.41556 &	0 &	0.83415 \\
            &         &$\alpha=90$, $\beta=73.1325$, $\gamma=90$&
S &	4i &	0.33437 &	0.5 &	0.41585\\
           &         &     &H &	4i &	0.00217&	0.0&
           0.24756\\
 &  &    &H&	4i &	0.25205 &	0.5 &	0.99709\\\hline
\end{tabular*}
\end{center}
\label{tabHS2}
\end{table}

\subsection{Crystal structure of H$_3$S at 200 and 250 GPa}

\begin{figure}[h]
\includegraphics[width=0.4\textwidth]{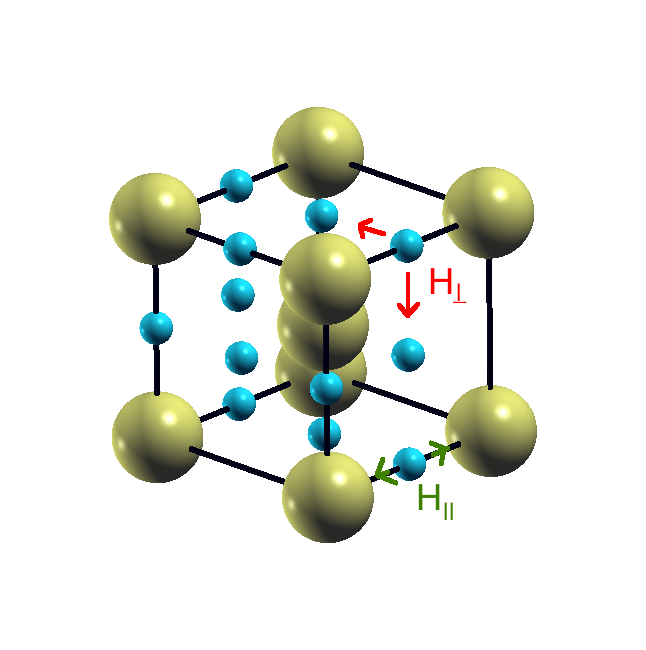}
\caption{Crystal structure of H$_3$S at 200 and 250 GPa. Hydrogen is
  depicted in cyan while sulfur is in yellow. The red (green) arrows
  label the H--S bond-bending (-stretching) modes, labeled H$_\perp$
  (H$_\parallel$). The volumes at 200 and 250 are 13.3334 \AA$^3$, and
  12.4925 \AA$^3$, respectively.}
\label{free_elec_bands}
\end{figure}

\begin{figure}[h]
\includegraphics[width=0.4\textwidth]{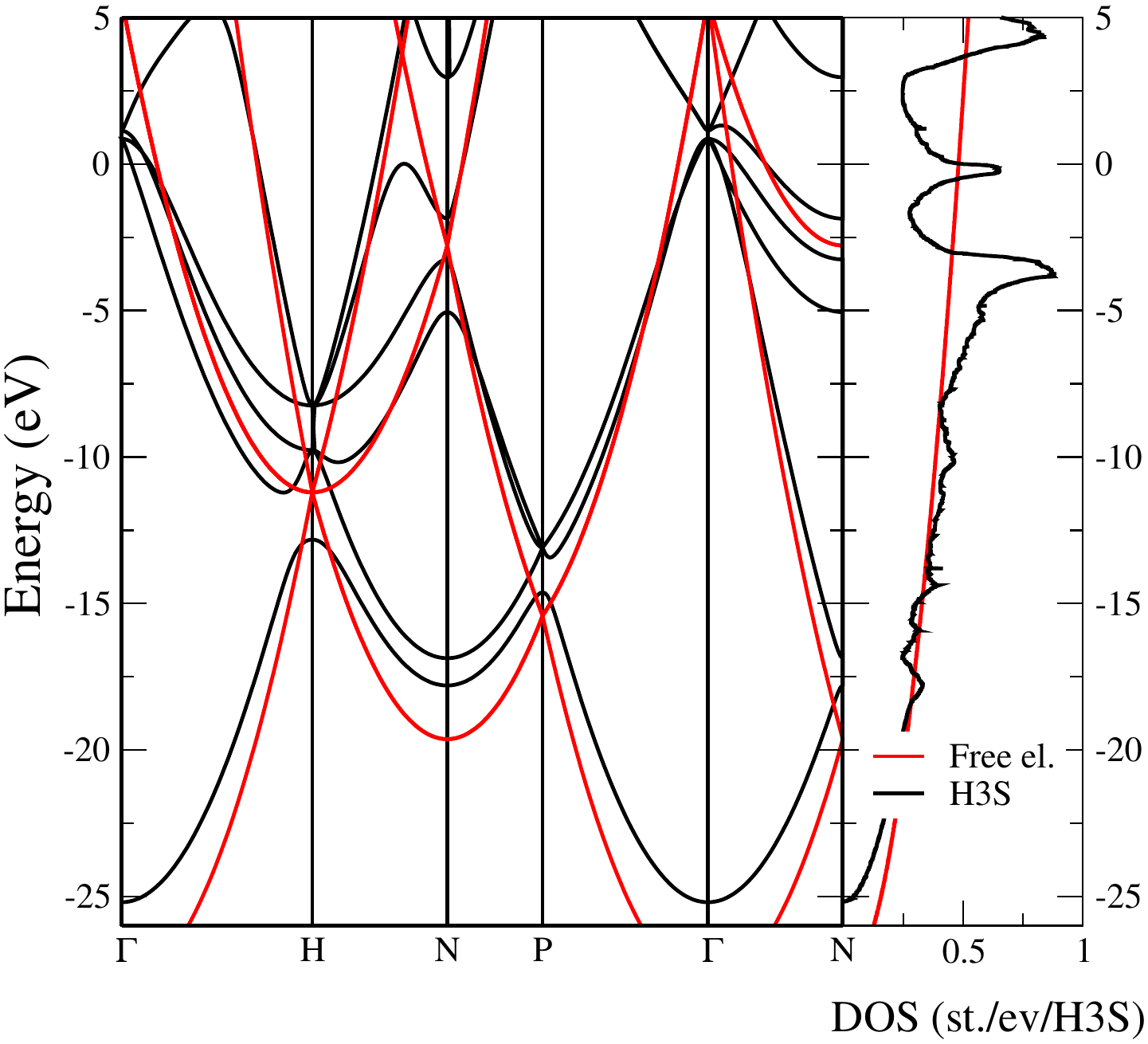}
\includegraphics[width=0.4\textwidth]{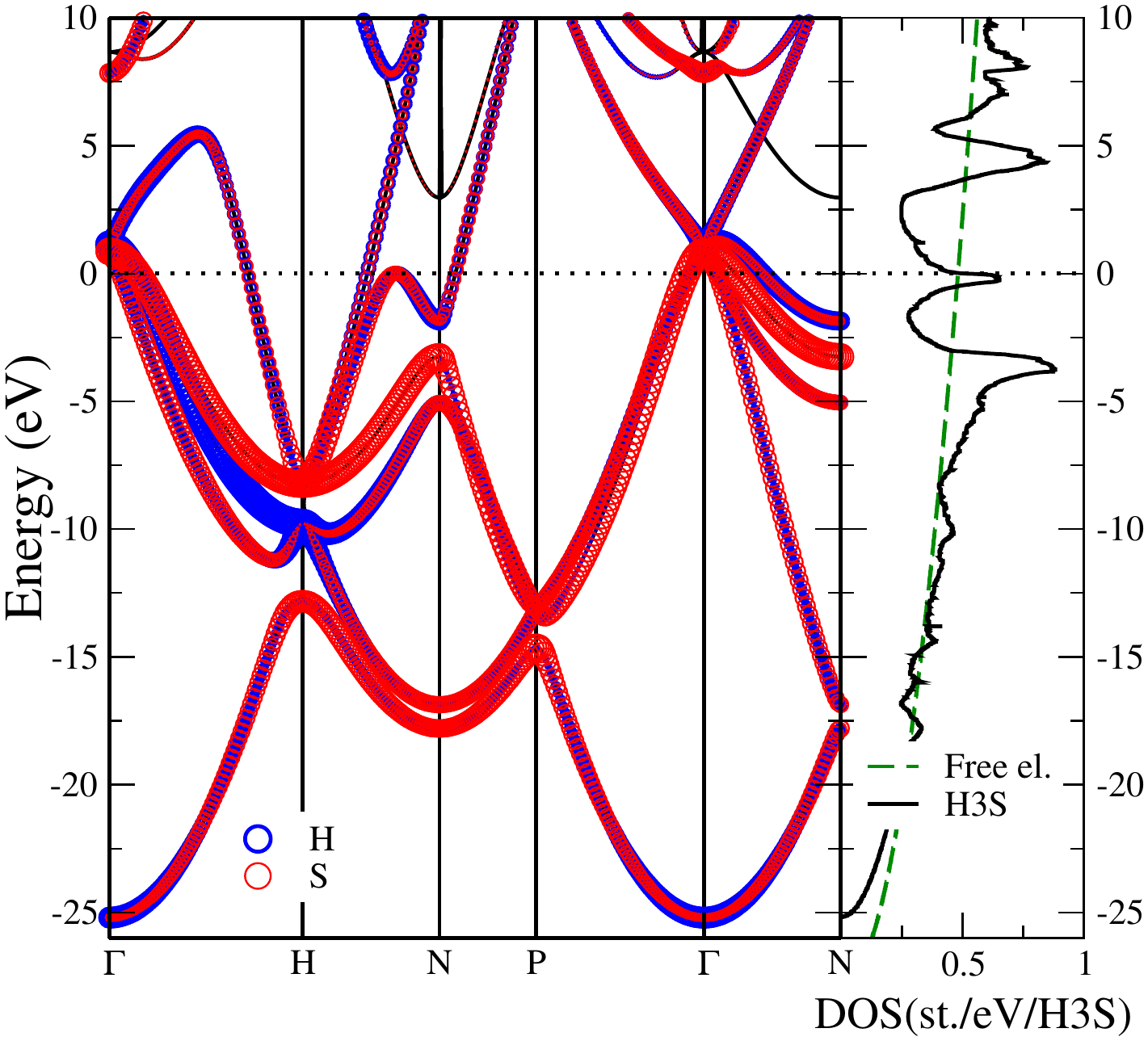}
\caption{Left: Free-electron bandstructure and density of states
  calculated for a bcc lattice with the same lattice parameter as
  H$_3$S at 200 GPa. The electronic structure and density of states of
  H$_3$S at 200 GPa are also shown for comparison.  Right: Electronic
  structure of H$_3$S. The thickness of the band is proportional to
  the projection of the electronic state over a chosen atomic orbital
  (fat bands representation).  }
\label{free_elec_bands}
\end{figure}

\subsection{Electronic structure of H$_3$S at 200 GPa}

The electronic structure of H$_3$S at $200$ GPa is shown in Fig.\
\ref{free_elec_bands} in the {\it fat-bands} representation including
decomposition into H and S atomic states.  In a 20 eV energy window
around the Fermi level ($\epsilon_f$), the electronic structure can be
fairly well interpreted in terms of free electrons on a bcc lattice.
However, in the proximity of the Fermi level, there is a substantial
hybridization between the H and S electronic states, leading to
avoided crossings at special points N, $\Gamma$ and along the H-N high
symmetry direction. The hybridization results in a peak in the density
of states at approximately $0.17$ eV below $\epsilon_f$.  At precisely
$\epsilon_f$, the density of states per spin is $N(0)=0.33$
states/eV/spin/H$_3$S cell, which is essentially identical to the
free-electron value.

\begin{figure}[h]
\includegraphics[width=0.4\textwidth]{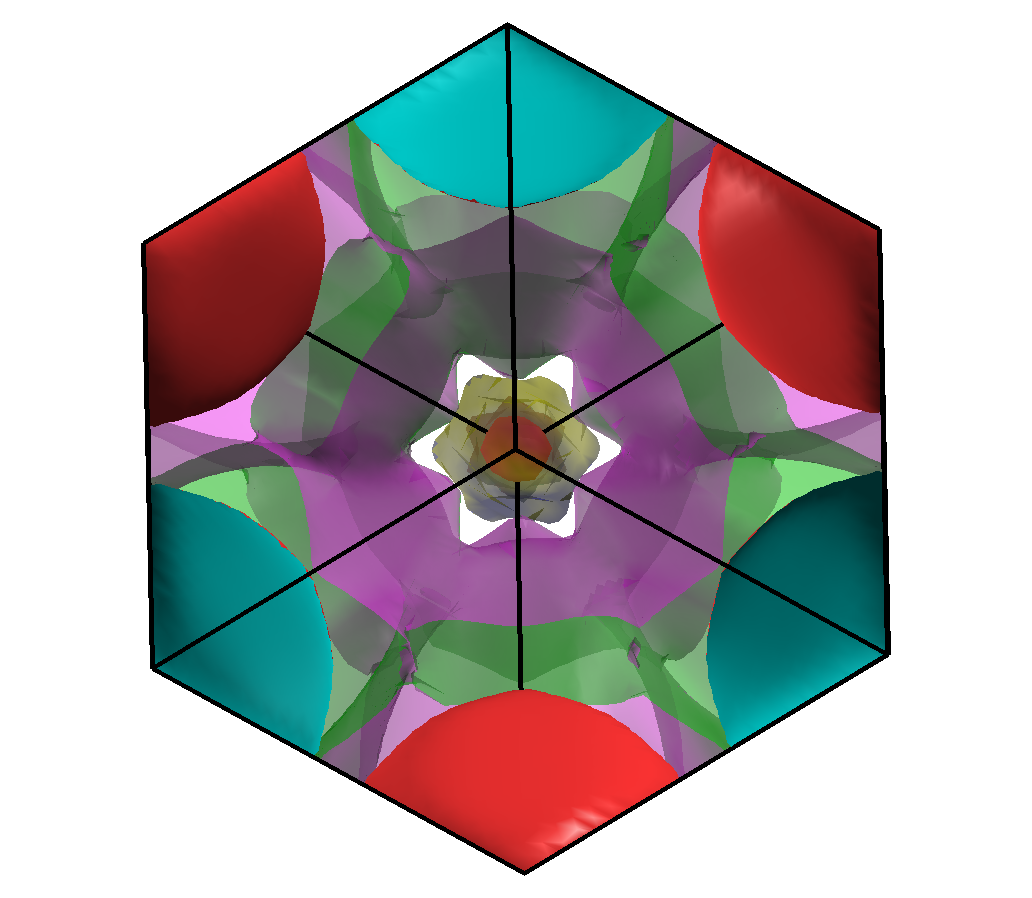}
\caption{Fermi surface of H$_3$S at 200 GPa including 5 sheets
  (right). The hole pockets at $\Gamma$ are visible in the center. The
  large Fermi surface contributing most of the density of states is
  shown in transparent colors.}
\label{free_elec_bands}
\end{figure}

The Fermi surface is composed of 5 sheets. The avoided crossing at the
zone center generates three hole pockets centered at $\Gamma$ and a
cubical electron Fermi surface centered at the N point. Finally, an
additional large Fermi surface sheet arises from the electron-pocket
at H.

\subsection{Vibrational and superconducting properties of D$_3$S at
  200 GPa}
\begin{figure}[h!]
\includegraphics[width=0.46\textwidth]{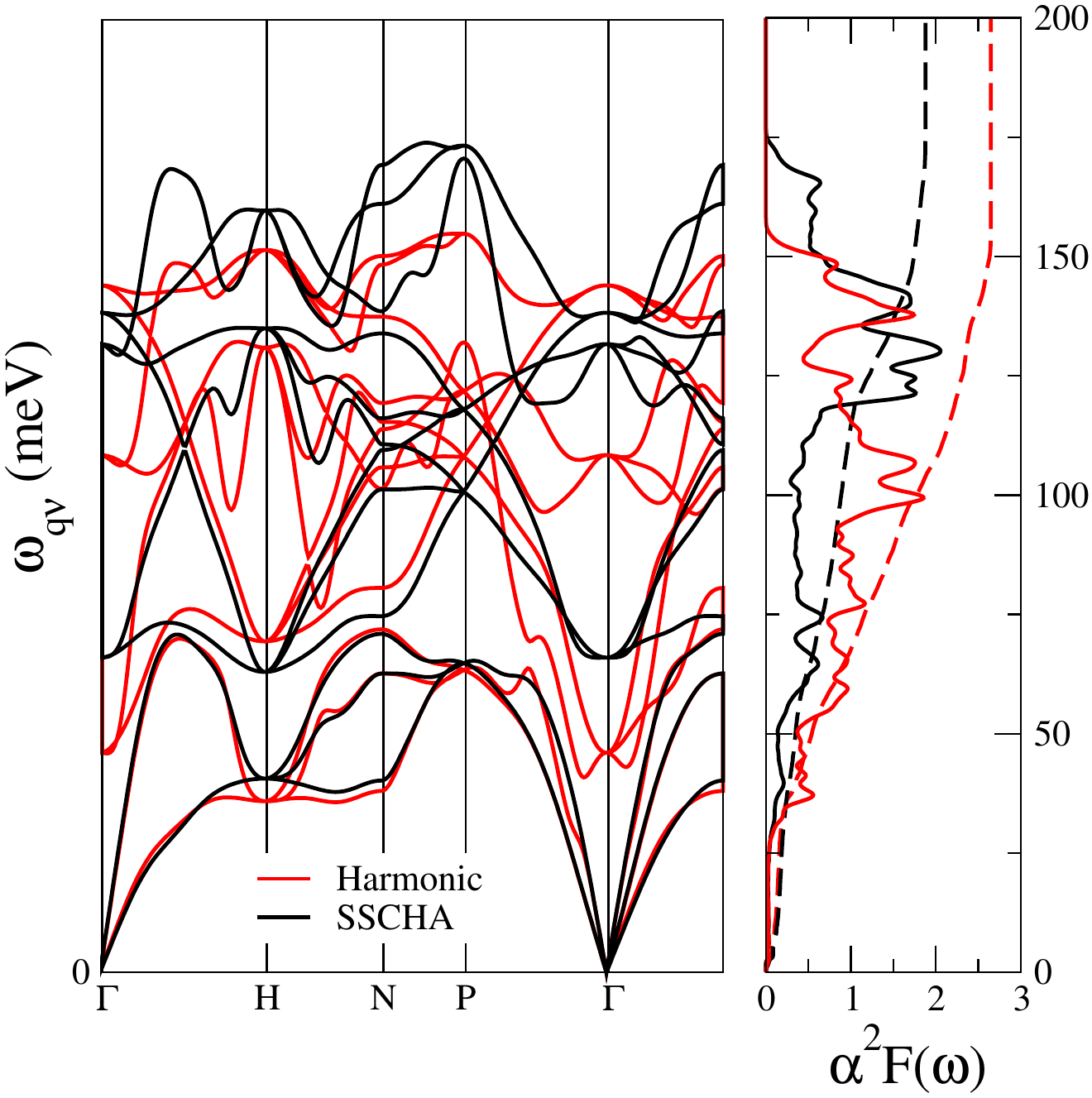}
\caption{Phonon spectrum, Eliashberg function and integrated
  electron-phonon coupling of D$_3$S at the harmonic and anharmonic
  levels (SSCHA).}
\label{D3S}
\end{figure}

The phonon dispersion of D$_3$S calculated using harmonic linear
response theory and the stochastic self-consistent harmonic
approximation is shown in Fig.\ \ref{D3S}.

\subsection{Vibrational and superconducting properties of H$_3$S at
  250 GPa}

\begin{figure}[h!]
\includegraphics[width=0.46\textwidth]{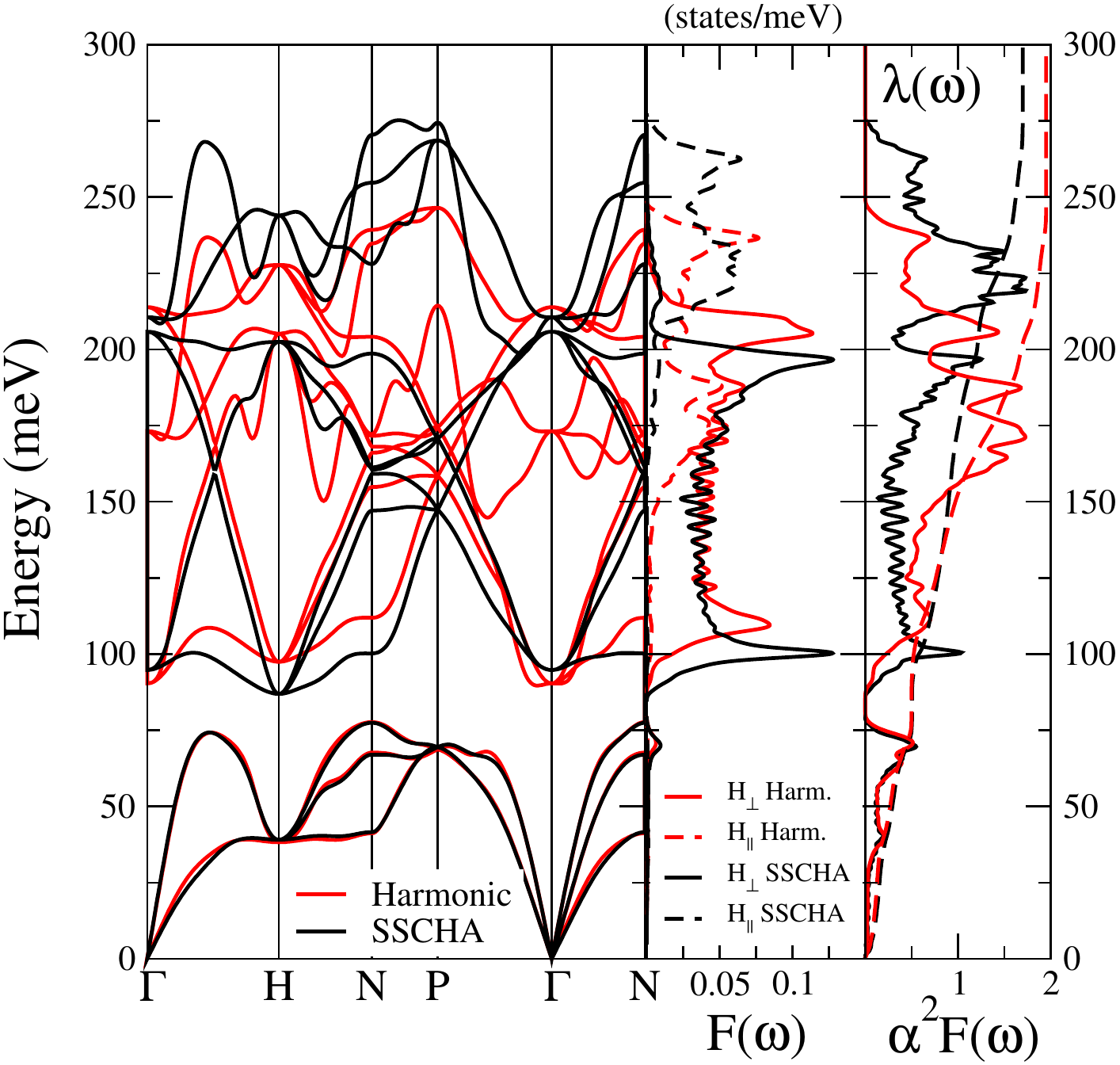}
\caption{Phonon spectrum, phonon density of states projected onto
  selected vibrations and the Eliashberg function and integrated
  electron-phonon coupling of H$_3$S at $250$ GPa  at the harmonic and anharmonic
  levels (SSCHA).}
\label{H3S250}
\end{figure}

The phonon dispersion of D$_3$S calculated using harmonic linear
response theory and the stochastic self-consistent harmonic
approximation are shown in Fig.\ \ref{H3S250}.

\subsection{Migdal-Eliashberg}

\begin{figure}[h]
\includegraphics[width=0.46\textwidth]{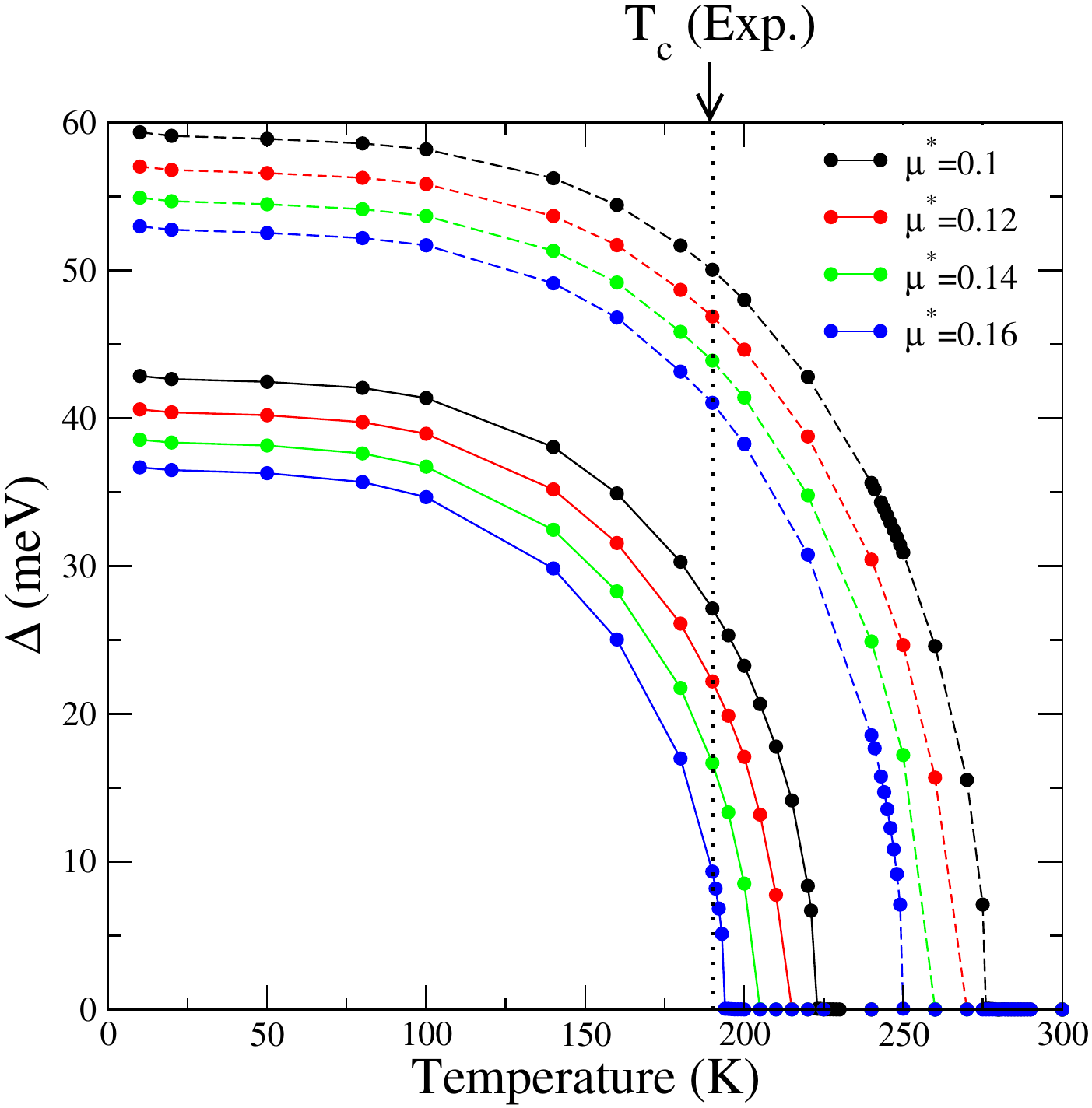}
\caption{Superconducting gap ($\Delta$) as a function of temperature
  from the solution of the isotropic Migdal-Eliashberg equations
  applied to H$_3$S.  The continuous lines refer to anharmonic phonon
  (SSCHA) while the harmonic phonon are denoted by dashed lines.}
\label{figSCGap}
\end{figure}

We solve the Isotropic Migdal-Eliashberg (ME) equations using either
the harmonic Eliashberg function or the one calculated within the
SSCHA. The equations are solved in the Matsubara
frequency space using $512$ Matsubara frequencies. The superconducting
gap is obtained from the lowest Matsubara gap $\Delta_{n=0}=\Delta$
and is plotted in Fig.\ \ref{figSCGap}.

\subsection{Superconducting properties using different approximations}

\begin{table*}[hb!]
  \caption{Electron-phonon interaction and a logarithmic average of phonon
    frequencies using the SSCHA, and without anharmonic effects. The
    T$_c$s are calculated using the SSCHA phonon spectrum and with 
    either the McMillan equation (T$_C^{\rm MM}$) or by solving the 
    isotropic Migdal-Eliashberg equations (T$_c^{\rm ME}$).
    A value of $\mu^*=0.16$ is used.} 
\squeezetable
\begin{tabular*}{\textwidth}{l c c c c c c c c c } 
\hline
\hline
Compound & $\lambda^{\rm har} $ & $\omega_{\rm log}^{\rm har}$ (meV) &
$\lambda^{\rm anh}$ & $\omega_{\log}^{\rm anh}$ (meV) & $T_c^{\rm MM,
 har}$ & 
$T_c^{\rm MM, anh}$ &T$_c^{\rm ME, har}$ &  T$_c^{\rm ME, anh}$ & T$_c$(Exp)\\
\hline
H$_3$S (200 GPa) & 2.64 & 90.4  & 1.84  & 92.86 & 158.8 & 124.9  & 250 &
194.0 &  190\\
H$_3$S (250 GPa) & 1.96 & 109.1  & 1.71  & 101.3 & 155.25 &   127.2 & 226 & 190 &  \\
D$_3$S (200 GPa) & 2.64 & 68.5 &  1.87  & 73.3  & 120.4 & 100.3 & 183 &
152.0 &  90\\
\end{tabular*}
\label{table1}
\end{table*}

\newpage

\subsection{Effects of the vibrational energy on the pressure}
\label{sec-zpe}

\begin{figure}[h]
\includegraphics[width=0.8\textwidth]{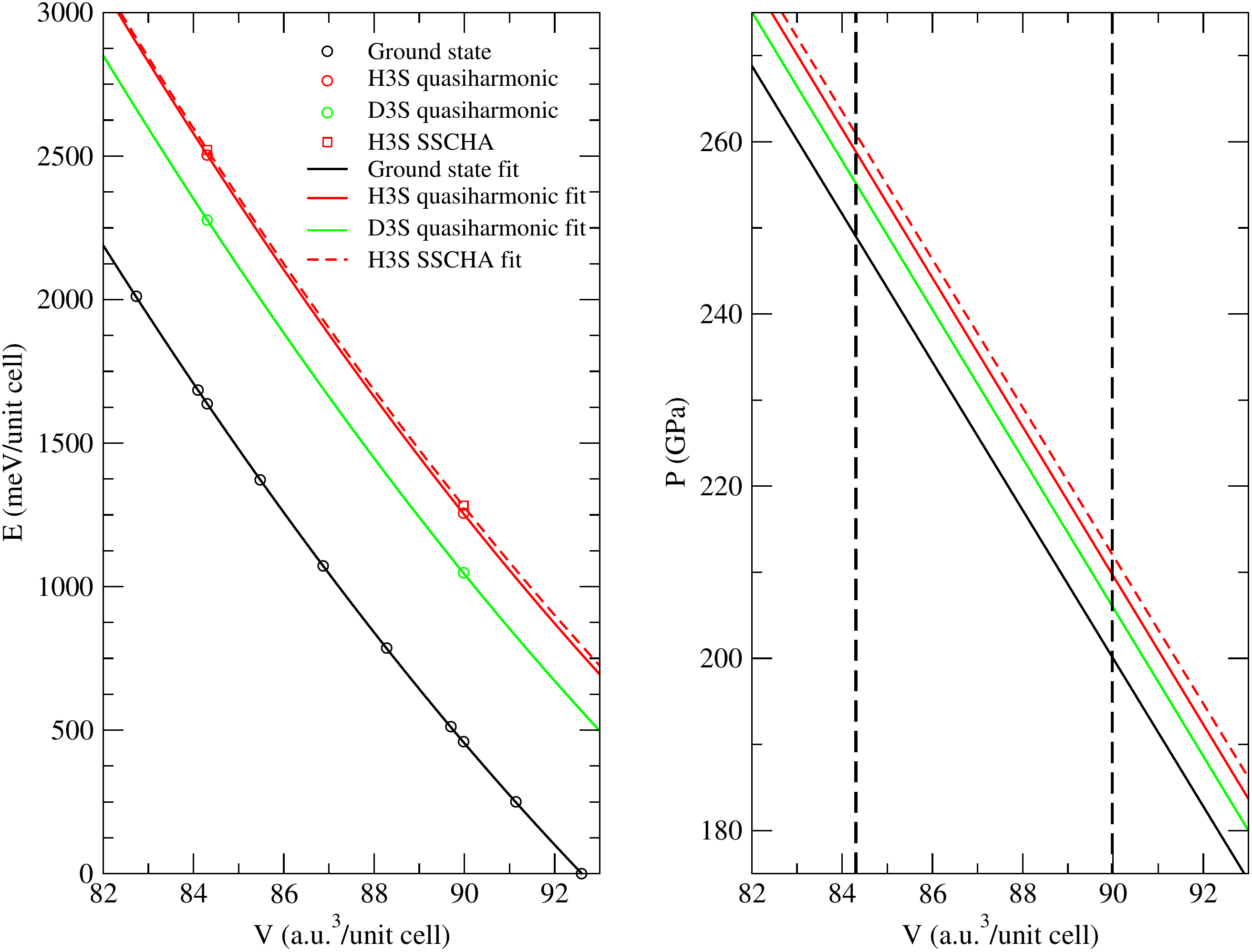}
\caption{(Left panel) Ground state energy without ZPE, including ZPE
  at the quasiharmonic level, and at the SSCHA level. The lines
  represent fitted curves following the recipe described in Sec.\
  \ref{sec-zpe}.  (Right panel) The pressure derived from the fitted
  energy curves. The vertical dashed lines denote the volumes used in
  the SSCHA calculations.}
\label{thermo}
\end{figure}

\begin{figure}[h]
\includegraphics[width=0.5\textwidth]{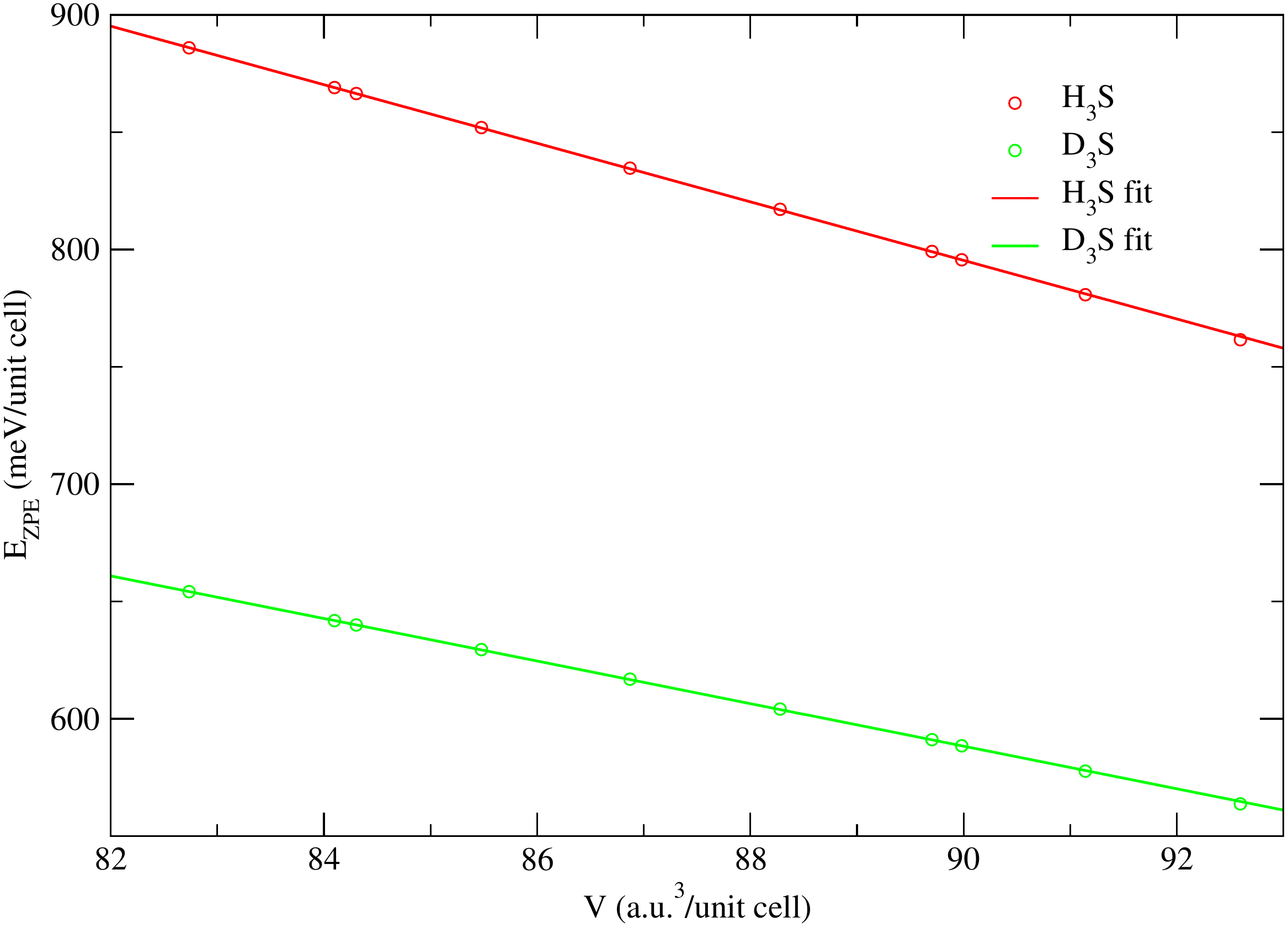}
\caption{ZPE calculated within the quasiharmonic level for H$_3$S and
  D$_3$S.  The solid lines denote the linear fit obtained with two
  data points, the volumes are those from the right panel of Fig.\
  \ref{thermo}.}
\label{zpe}
\end{figure}

In Fig.\ \ref{thermo} the contribution of the atomic vibrations to the
total energy is shown in the quasiharmonic approximation and
SSCHA. The ground state total energy, without the zero point energy
(ZPE), is fitted to a second order polynomial as a function of the
volume $V$, $E_0(V) = A_0 + B_0 V + C_0 V^2$.  The fitting parameters
are reported in Table \ref{t-zpe}. We add the vibrational contribution
to $E_0(V)$ calculated in the quasiharmonic approximation and in the
SSCHA.  This contribution is calculated at two volumes, those shown in
the right panel of Fig.\ \ref{thermo}. The vibrational energy $E_v$ is
fitted linearly to $E_v (V) = A_v + B_v V$. As shown in Fig.\
\ref{zpe}, the linear fit provides a very good approximation to $E_v
(V)$. The linear form of $E_v (V)$ is obtained within the
quasiharmonic approximation for H$_3$S and D$_3$S, and in the SSCHA
for H$_3$S. The ZPE energy is obtained using a 6$\times$6$\times$6
phonon mesh.

The correction to the pressure from including the atomic vibrations is
calculated from the $E(V) = E_0 (V) + E_v (V)$ curves and $P(V) = -
\mathrm{d} E(V) / \mathrm{d} V$.  These curves are shown in Fig.\
\ref{thermo}. Neglecting the ZPE underestimates the pressure. The
correction to the pressure from the ZPE is smaller for D$_3$S than for
H$_3$S due to the smaller ZPE of deuterium. The SSCHA gives a small
correction to the pressure obtained within the quasiharmonic
approximation. The pressure corrections to the volumes for which the
electron-phonon calculations were performed are summarized in Table
\ref{t-p}.

\begin{table}[b]
  \caption{Calculated parameters for the quadratic fit of $E_0(V)$ and the linear fit of $E_v(V)$.}
\begin{center}
\begin{tabular*}{0.69\textwidth}{c c c c}
\hline
\hline
          & $A_0$ (mev) & $B_0$ (meV/a.u.$^3$) & $C_0$ (meV/a.u.$^6$) \\
\hline
          & 49319       & -900.88            & 3.9772 \\
\hline
          & $A_{v}$ (mev) & $B_{v}$ (meV/a.u.$^3$) & \\
\hline
H$_3$S quasiharmonic & 1918.5    & -12.479            & \\
D$_3$S quasiharmonic & 1404.7    & -9.0707            & \\
H$_3$S SSCHA         & 1814.2    & -11.027            & \\
\hline
\hline
\end{tabular*}
\end{center}
\label{t-zpe}
\end{table}

\begin{table}[b]
\caption{Correction to the pressure $P$ from the vibrational energy.
}
\begin{center}
\begin{tabular*}{0.75\textwidth}{c | c | c | c | c}
\hline
\hline
 $V$ (a.u.$^3$) & \multicolumn{4}{c}{$P$ (GPa)} \\ 
              & no ZPE & D$_3$S quasiharmonic & H$_3$S quasiharmonic & H$_3$S SSCHA \\
\hline
89.9822       & 200       & 206      & 210   & 212 \\
84.3035       & 250       & 255      & 259   & 261 \\
\hline
\hline
\end{tabular*}
\end{center}
\label{t-p}
\end{table}


\begin{thebibliography}{99}
\bibitem{TecDet} Our results were obtained from first-principles DFT
  linear-response calculations as implemented in the
  QUANTUM-ESPRESSO~\cite{QE} package.  We used ultrasoft \cite{US}
  pseudopotentials, a generalized gradient approximation \cite{PBE}, a
  plane-wave cutoff energy of 60 Ry on the kinetic energy and 600 Ry
  on the charge density.  The charge density and dynamical matrices
  were calculated using a 32$^3$ Monkhorst-Pack shifted
  electron-momentum grid and an Hermitian-Gaussian smearing of 0.03
  Ry.  The average electron-phonon coupling was obtained using Wannier
  interpolation \cite{CalandraWannier}.  We used a 32$^3$
  electron-momentum grid randomly shifted from the origin and a 32$^3$
  Monkhorst-Pack shifted phonon-momentum grid with a smearing of
  $0.03$ eV.  The phonon linewidth at a given phonon momentum was
  calculated using a 50$\times$50$\times$50 electron-momentum mesh
  (randomly shifted from the origin).
\bibitem{footnote} The pressure computed here is without inclusion of
  ZPE. Neglecting ZPE underestimates the pressure of $\approx$ 12 GPa
for H$_3$S and $6$ GPa for D$_3$D for a volume of 89.9822 (a.u.)$^3$.
See \cite{SI} for more details.
\bibitem{SSCHAtechdet} The SSCHA calculations were performed using a
  3x3x3 supercell for both H$_3$S and D$_3$S at 0 K, yielding
  dynamical matrices on a commensurate 3x3x3 $q$-point grid.  An
  additional calculation at 200 K conformed that temperatures within
  the range of the predicted superconductivity does not affect the
  phonon spectra.  The difference between the harmonic and anharmonic
  dynamical matrices in the 3x3x3 phonon momentum grid was
  interpolated to a 6x6x6 grid.  Adding the harmonic matrices to the
  result, the anharmonic dynamical matrices were obtained in a 6x6x6
  grid.  These dynamical matrices were used for the anharmonic
  electron-phonon coupling calculation.
\end{thebibliography}
\end{document}